# Star Maps and Travelling to Ceremonies – the Euahlayi People and Their Use of the Night Sky

Robert S. Fuller[1,2], Michelle Trudgett[1], Ray P. Norris[1,4], Michael G. Anderson[3]

[1]Warawara, Department of Indigenous Studies, Macquarie University NSW 2109, Australia
Email: *robert.fuller1@students.mq.edu.au, michelle.trudgett@mq.edu.au*
[2]Macquarie University Research Centre for Astronomy, Astrophysics and Astrophotonics, Macquarie University NSW 2109, Australia
[3]Euahlayi Law Man, PO Box 55, Goodooga, NSW, 2838, Australia
Email: *ghillar29@gmail.com*
[4]CSIRO Astronomy and Space Science, PO Box 76, Epping, NSW, 1710, Australia
Email: *raypnorris@gmail.com*

## Abstract

The Euahlayi people are an Australian Aboriginal language group located in north-central New South Wales and south-central Queensland. They have a rich culture of astronomy, and use of the night sky in resource management. Like several other Aboriginal peoples, they did not travel extensively at night, and so were assumed not to use the night sky for navigation. This study has confirmed that they, like most other Aboriginal groups, travelled extensively outside their own country for purposes of trade and ceremonies. We also found that, previously unknown, they used "star maps" in the night sky for learning and remembering waypoints along their routes of travel, but not for actual navigation. Further research may find that this was common to many Aboriginal groups in Australia.

## Notice to Aboriginal and Torres Strait Islander Readers

This paper contains the names of people who have passed away.

## 1. Introduction

Cultural astronomy is defined as the study of the effect of astronomical knowledge or theories on ideologies or human behaviour (Campion, 2003: xv). Fuller et al (2014a: 3-4) report that while there is a rich knowledge of Aboriginal astronomy, the literature on Kamilaroi and Euahlayi astronomy, based on ethnography from the 19[the] century, is limited and contains many contradictions. They collected knowledge of the sky from current Kamilaroi, Euahlayi, and neighbouring communities, including many interviews and recordings of stories during 2013 from eight participants from those communities. Those participants, referred to in the texts as P1……P8, are noted in the Acknowledgements. One participant, Michael Anderson, with both Euahlayi and Kamilaroi heritage, provided such a complete description of star maps as it related to his culture that he is included as an author of this paper.

Fuller et al (2014a)) confirmed the hypothesis that the knowledge gained could add to the current body of knowledge of Australian Aboriginal sky culture. Most of the data was released under the terms of Ethics Approval 5201200462 by Macquarie University. This paper presents previously unpublished data from the study used to determine whether the knowledge about star maps collected through the larger project adds a deeper level of understanding into the sky culture of the Kamilaroi and the Euahlayi peoples.

The stories collected by Fuller et al (2014a) do not just entertain and describe some physical object in the sky. Aboriginal culture is oral in nature, and oral transmission of knowledge is extremely important, particularly in regards to Law. As Aboriginal Law governs all aspects of Aboriginal life, it establishes a person's rights and responsibilities to others, the land, and natural resources (Law Reform Commission of WA, 2006: 64). Cultural stories transmit Law, and in this respect may have different levels of meaning. Sveiby and Skuthorpe (2006: 45-51) described four levels; one being for children (to explain nature), others being for relationships between people, relationships between the community and country, and ceremonial practices. A participant in the project said that some stories could have up to "30 levels" of meaning. While we have avoided reference to the ceremonial levels of meaning in this study, in the case of the use of the sky for travelling to ceremonies, we may reference ceremonial matters where permission has been received from the owners of such ceremonial knowledge.

## 2. The Euahlayi People

The Euahlayi people are an Australian Aboriginal cultural group located in the northwest of New South Wales (NSW). Ash et al (2003:1) have described the Euahlayi language as "Yuwaalayaay", but Euahlayi participants in the project maintain that "Yuwaalayaay" is a clan language of the neighbouring Kamilaroi people, and that Euahlayi is a separate language (P2). Mathews (1902: 137) described the "Yualeai" language as:

> *The natives speaking this language are located upon a tract of country in southern Queensland, including the Bokhara, Birrie, Narran, Ballonne and Moonie Rivers, and extend some distance within the New South Wales frontier, where they are met by the Kamilaroi nation.*

When this area was described by Sim and Giacon (1998: vii and xii), Sim described the language in the area as "Yualeiai", while Giacon (the Editor) changed this to Yuwaalayaay, so there have been a number of interpretations of the name over the years. Fig 1 shows the area of the Euahlayi language group (labelled "Juaaleiai").

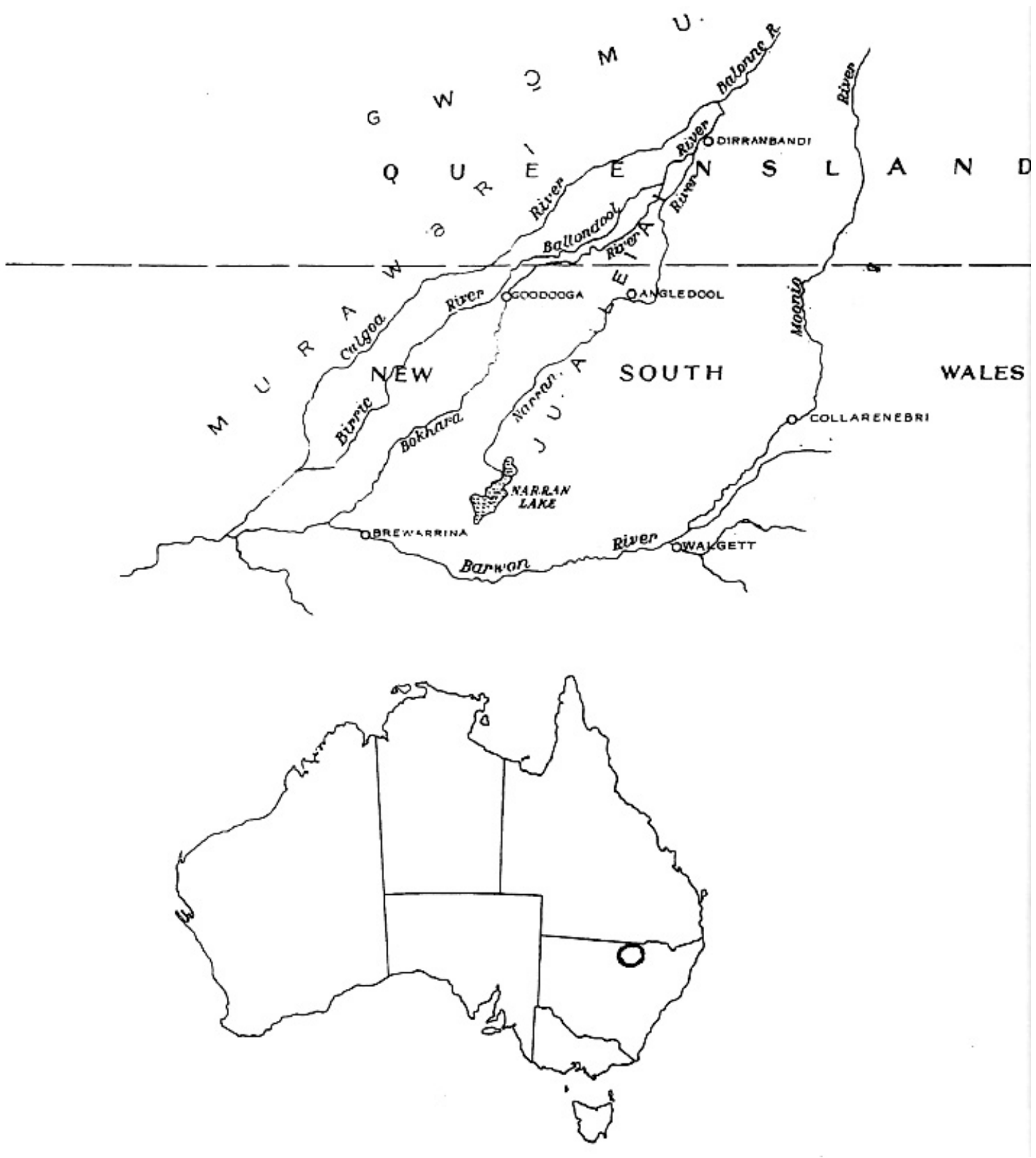

*Fig 1 location of the Euahlayi language group (Sim & Giacon 1998)*

DNA research suggests that Aboriginal Australians are descendants of people who left the Middle East approximately 70,000 years ago (Rasmussen et al, 2011: 98). Their ancestors may have settled in southeast Australia as long as 40,000 years ago, based on radiocarbon dating of the Mungo Man burial at the Willandra Lakes region of NSW (Bowler et al, 2003: 840).

The population of the Kamilaroi and Euahlayi cultural grouping was estimated at 15,000 in 1788 although participants point out that the resources available could maintain a population as large as 60,000. This may have dropped to as low as 1000 in 1842 (Sveiby & Skuthorpe, 2006: 25-6). As a result of pressure from European colonisers, there was a movement of Aboriginal people in this group towards the northwest. The current population of people identifying as Kamilaroi and/or Euahlayi ancestry is approximately 29,000 (estimates from Kamilaroi Nation Applicant Board). The same source estimates the current number of people with Euahlayi ancestry as being “around 3000”, implying that the population at the time of European invasion was small.

## 3. Aboriginal travel and navigation using the stars

### 3.1 Navigation at night

Haynes and Haynes (1996: 7-8) described the difference between Aboriginal and European astronomy:

> *Astronomy, in the sense of a comprehensive and coherent body of knowledge about the stars, was an integral component of Aboriginal culture. Like the*

> *Newtonian-based system of Western science, it represented an attempt to construct a view of the Universe as an ordered and unified system, but in most ways it was fundamentally different. It was relational rather than mathematically-based, and it was concerned with similarity rather than with difference, with synthesis rather than analysis, with symbiosis rather than separation……..For the Aborigines, the stars not only evoked wonder, they predicted and explained natural occurrences and provided celestial parallels with tribal experiences and behavioural codes.*

A common question is: Do Aboriginal people use the stars for navigation at night? Norris and Harney (2014) show that Wardaman people prefer to travel at night, and use the stars for navigation. Maegraith (1932: 25) stated that Aboriginal people of the Central Desert were unable to navigate at night, notwithstanding their detailed knowledge of the night sky. Lewis (1976: 273-4), as a part of a project in Aboriginal route finding with Walbiri and Pintupi people in the Simpson Desert, found that his participants did not use the stars for navigation, and were disoriented at night when unable to see landscape features. The answer to this question therefore seems to vary from one language group to another.

One of the participants (P2) stated that while Aboriginal people had a clear understanding of the sky at night, and could use stars and other objects for directions, they had no interest or purpose in travelling at night.

Kerwin (2010: 68-9) lists anecdotal stories from participants in his research showing their ability to navigate in their country using the stars at night. Cairns & Harney (2003: 9) describe the Dreaming Track in the sky which helps navigation on the ground, and Norris and Harney (2014) give other examples.

There is much evidence in the literature about the Aboriginal use of the night sky for predicting when resources would be available. Fuller et al (2014b) have shown how the Emu in the Sky, an important object outlined by the dust clouds in the Milky Way, governed the Euahlayi people's timing of the use of the emu egg resource on the ground. Almost all the literature on Aboriginal astronomy refers to the central role of the night sky in Aboriginal cultural stories. Sveiby and Skuthorpe (2006: 41) describe Aboriginal storytelling as "dramatic art", and one can imagine that many of the stories collected since European invasion were told at night, using the key cultural figures and animals of the Dreaming that were connected to objects in the sky. These stories had many levels (ibid: 42-50), from children's stories to entertain or describe natural phenomena, to levels describing spiritual action and ceremonies.

## 3.2 Travel

Research on Aboriginal culture in the last two centuries has brought a change in the original view by Europeans that Aboriginal peoples were nomadic hunter-gatherers eking out a subsistence existence. As early as the 1840's Curr speculated that Aboriginal people in Victoria (VIC) had an excess of leisure time after fulfilling their needs (Sahlins, 1972: 24). Sahlins (ibid: 9-32) coined the description "The Original Affluent Society" for hunters and gatherers, including Australian Aboriginal peoples, and used data from early ethnographic studies in Australia to show that they were able to meet their needs with adequate time to spare for leisure, and that travel was limited to a need to access resources. Kerwin (2010) has made a strong case for travel by Aboriginal people being mainly for trade and ceremony, but that otherwise, most Aboriginal people living in the better resourced parts of Australia lived sedentary lives in clearly defined territories. He and earlier writers, such as McBryde (2000: 157-64), Mulvaney and Kamminga (1999: 95), and Flood (1983: 235-6) have described a vast network of trade routes throughout Australia, trading in goods, ceremonies, and stories. These included well-established trade routes connecting the inland with coastal Queensland (QLD), and down the Dividing Range to the Snowy Mountains. Commodities traded included bunya nuts, pituri (a nicotine- based narcotic), stone axes, ochre, and wooden implements. Stories and ceremonies were also traded along the same routes, and people from different language groups attended and participated in ceremonies such as the *Bora* initiation ceremony of the southeast of Australia. Reynolds (1981: 11-2) described the "large ceremonial gatherings (that) provided the venue for gossip, trade and cultural interchange".

### 3.3 Songlines and Dreaming tracks

It is no coincidence that stories and ceremonies were traded along the trade routes of Aboriginal Australia. Kerwin (2010: 113-20) and McBryde (2000: 157-64) have shown that trade routes were equally important as songlines and Dreaming tracks. In some cases stories were also traded along trade routes, and eventually that trade route became a storyline or Dreaming track, and the route was incorporated into the story. For example, the Two Dog Dreaming story (Kerwin 2010: 37, 90) describes how the ancestral emu, Kuringii, was chased by two ancestral dingoes along a trade route from Cape York to South Australia (SA) through the Queensland Channel Country. Kuringii was eventually killed at the foot of the Flinders Ranges, and his blood is the source of the prized red ochre from Parachilna in SA. In this way, the ochre traded on this route became part of the story in such a way that the route is now a songline for the song that relates the story. A Dreaming track is another way to express the meaning of a story which is a part of a traditional Dreaming story that travels down a trade route. Kerwin (2010: 113-4) says that a Dreaming story or songline can change from one end of the track to the other, but still has the same basic theme. The language of the story will also change as it moves through different language groups along the track, but as the story is sung, the "melody" remains the same; only the language changes. A person can recognise the story without understanding the language (P2). Kerwin (2010) also says that as the story travels over the landscape it will have changes relating to the local country of those singing the story, but the original theme remains the same.

Songlines can also be seen in the sky at night. Euahlayi people (P2) know of a songline stretching from Heavitree Gap at Alice Springs to Byron Bay on the East Coast. This is the songline of *Mulliyan-ga* (the eaglehawk), and runs from the star Achernar in the West overhead to Canopus, to Sirius, and then to the East. *Mulliyan-ga* fought, and was defeated by the caterpillar, *Yipirinya*, at Alice Springs after travelling from the East, and his spirit remains in Achernar. This is a songline which connects the Arrernte people of Alice Springs with the Euahlayi people of northwest NSW, and it is possible the Arrernte have a story which connects to *Mulliyan-ga*. The Euahlayi (ibid) also know of the Black Snake/Bogong Moth songline which connects Normanton on the Gulf of Carpentaria with the Snowy Mountains near Canberra. This songline in the sky follows the Milky Way, and intersects with the *Mulliyan-ga* songline over Euahlayi country.

## 4. Star maps and travelling to ceremonies

Cairns (1996, 2005) speculated that small depressions in the sandstone, known as cupules, at the Elvina Track site north of Sydney might represent maps of the stars in the sky. However, Bednarik (2008), an expert in rock art, has been very sceptical that cupules are anything other than natural phenomena in all but very limited cases.

Another use of the term “star map” describes the use of patterns of stars to represent routes of travel on land. During the summer months, Aboriginal people travelled through their own country, and often the country of other clans and language groups, to trade in goods and stories, and in particular, to attend and participate in ceremonies. These ceremonies often took place at special sites, such as *Bora* grounds, that had been used for such purposes for long periods of time, and usually marked a story or event of spiritual significance that took place at that location during the Dreaming. Early ethnographers documented these ceremonies in sufficient detail that we are able to find examples of ceremonies where it was clear that people from a wide area attended. Mathews (1894: 106-9) describes a *Bora* ceremony at Gundabloui (near Collarenebri NSW) in 1894 where the attendees travelled up to 160 km on foot to attend. The camp was broken down into three sections, one being people from the Mogil Mogil, Collarenebri, and Walgett areas (most likely Euahlayi language group), another from the Kunopia, Mungindi, and Welltown areas, and the last from the Moonie and St. George (QLD) areas. The latter two groups were possibly Kamilaroi and Bigambul language groups. According to Mathews, messengers were sent out after the *Bora* ceremony at Kunopia (near Mungindi NSW) two years prior to invite people to the Gundabloui *Bora*. As the time for the *Bora* approached, the people attending commenced travelling, with one of the messengers as a guide. They travelled by day and camped at night. In this case (ibid: 124), the ceremonies were held between 12 February and 10 March, but attendees began arriving a month before. *Bora* ceremonies took place over a wide range of months (Fuller et al, 2013: 31), but it appears that they were normally held during the summer months.

Fuller et al (2013) describe 68 known *Bora* grounds in the northwest and north central area of NSW and over the border in south central QLD, which consisted of a large circle of cleared earth or stone, connected by a cleared pathway to a smaller circle of cleared earth or stone. The larger circle was considered to be the “public” circle, where all participants could attend, and the smaller circle was the “sacred” circle, where only the initiates and elders were allowed. Mathews (1894: 106-9) suggested that Euahlayi people travelled long distances to attend *Bora* ceremonies so that people from individual clan groups would have had a need to know how to navigate through a very large area possibly using the Dreaming tracks described above. We are not aware of any actual portable “maps” having been found in Australia, and the only suggestions of way-finding devices have been stone or clay *“cyclons”* in NSW, *“toas”* (decorative objects in various materials from Central Australia), and message sticks, but the function of these objects for travel is not clear (Kerwin, 2010: 74-8).

The following description of the use of the sky has been provided by the author Anderson, and confirmed in part by the participants P4, P7, and P8. For the Euahlayi, and for the Kamilaroi and neighbouring language groups, there was another way to use the stars for travelling which was not a form of navigation by the stars. This was the use of patterns of stars (“star maps”) to teach people how to travel in and outside of their country. Knowledge in Aboriginal culture is transmitted orally, so this technique could be considered a memory aid to assist in teaching, and as a reminder for future travel. Initiated men, including the messengers mentioned by Mathews, would be the holders of this knowledge, but it is possible women were also included, as knowledge of travel was not necessarily ceremonial in nature. In Euahlayi country, the winter months of May, June, and July would be used for planning the travel to ceremonies during the summer months, starting as early as September. The people planning to travel would already know where they had been invited, as the messengers would have arrived with the invitations. A part of the early winter activities would be the travel plan, and at this time, young people (and perhaps women) would be taught how to travel using the songlines or Dreaming tracks described by the star maps. The knowledge holder would use a clear night at the right time of the year and point out the directions for travel, using the patterns of stars in the star map in the sky to guide the intended traveller from place to place on the ground using the stars as what we now call “waypoints” in terrestrial navigation. To the Aboriginal person, these waypoints could be a bend in a river, a waterhole, a marked tree, or a stone arrangement. Eventually the star map would lead to the destination, which would be the ceremonial ground.

An example, based on mid-May in the late evening, would be a star map leading to Carnarvon Gorge in QLD, which is a known ceremonial centre. This is a trip of over 600km. Looking at the southeast sky, the winter camp in Euahlayi country where the planners are located is represented as an area in the constellation Sagittarius bounded by the stars Epsilon Sagittarii, Beta Sagittarii, W Sagittarii, Delta Sagittarii, and the star cluster M7. This area would have also incorporated Kamilaroi and Murrawarri peoples. The star map to Carnarvon Gorge would proceed from the winter camp to the stars Gamma Scorpii (representing Dirranbandi, QLD), Kappa Scorpii (St. George, QLD), Theta Scorpii (Surat, QLD), Eta Scorpii (Roma, QLD), and Zeta Scorpii (Carnarvon Gorge). This is represented in Fig 2.

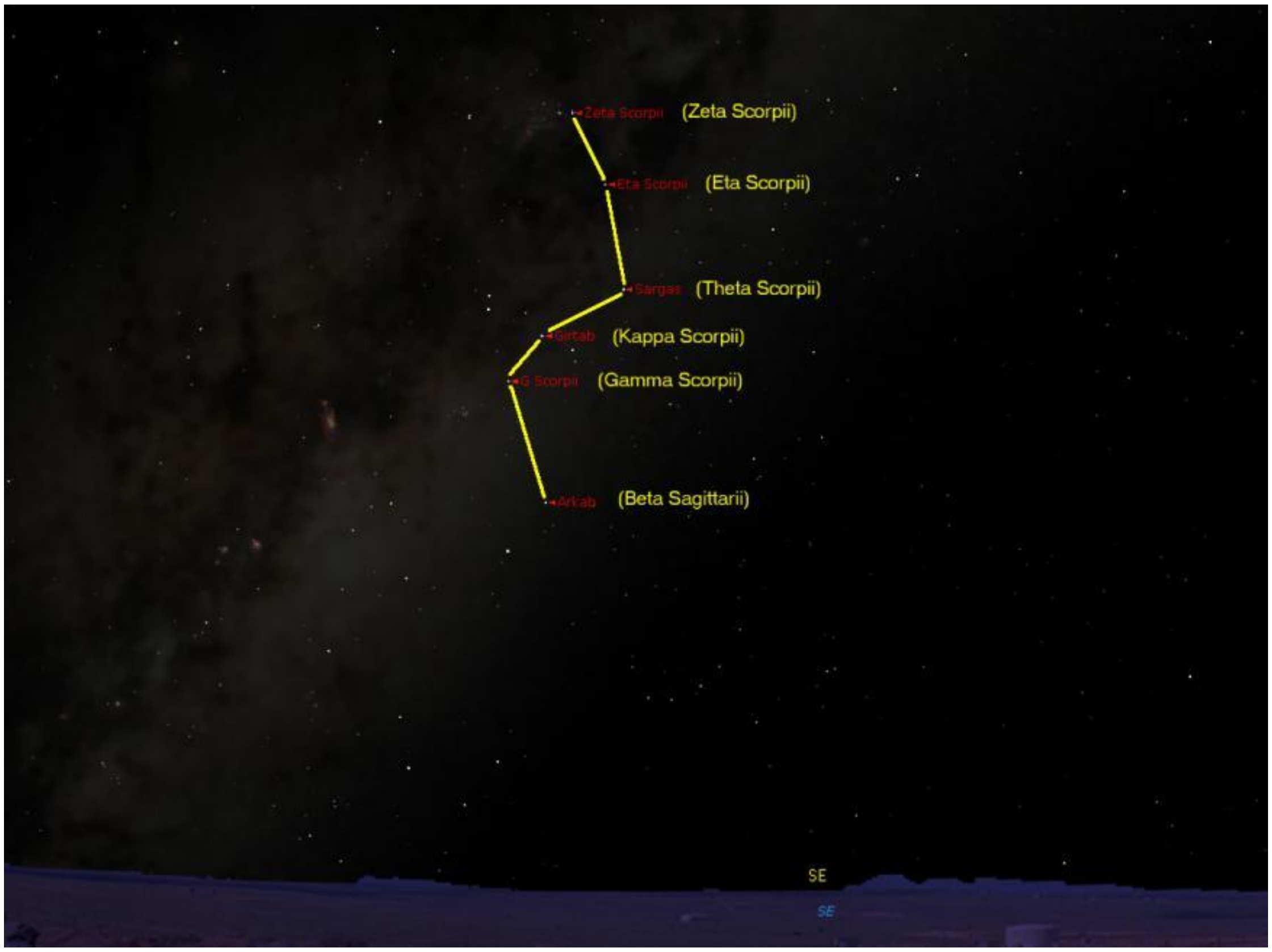

*Fig 2 Goodooga to Carnarvon Gorge star map (image courtesy of Starry Night Education)*

If the travel was to the Bunya Mountains (which was the source of the prized bunya nuts), the traveller would turn at Theta Scorpii (Surat, QLD), to Sigma Arae (Chinchilla, QLD), to Alpha Arae (Dalby, QLD), then to Beta and Gamma Arae (Bunya Mountains). This is represented in Fig 3. The travel to share in bunya nuts with the language groups whose country was the Bunya Mountains is well documented by Ridley (1875: 159) and Pietre (1904: 11).

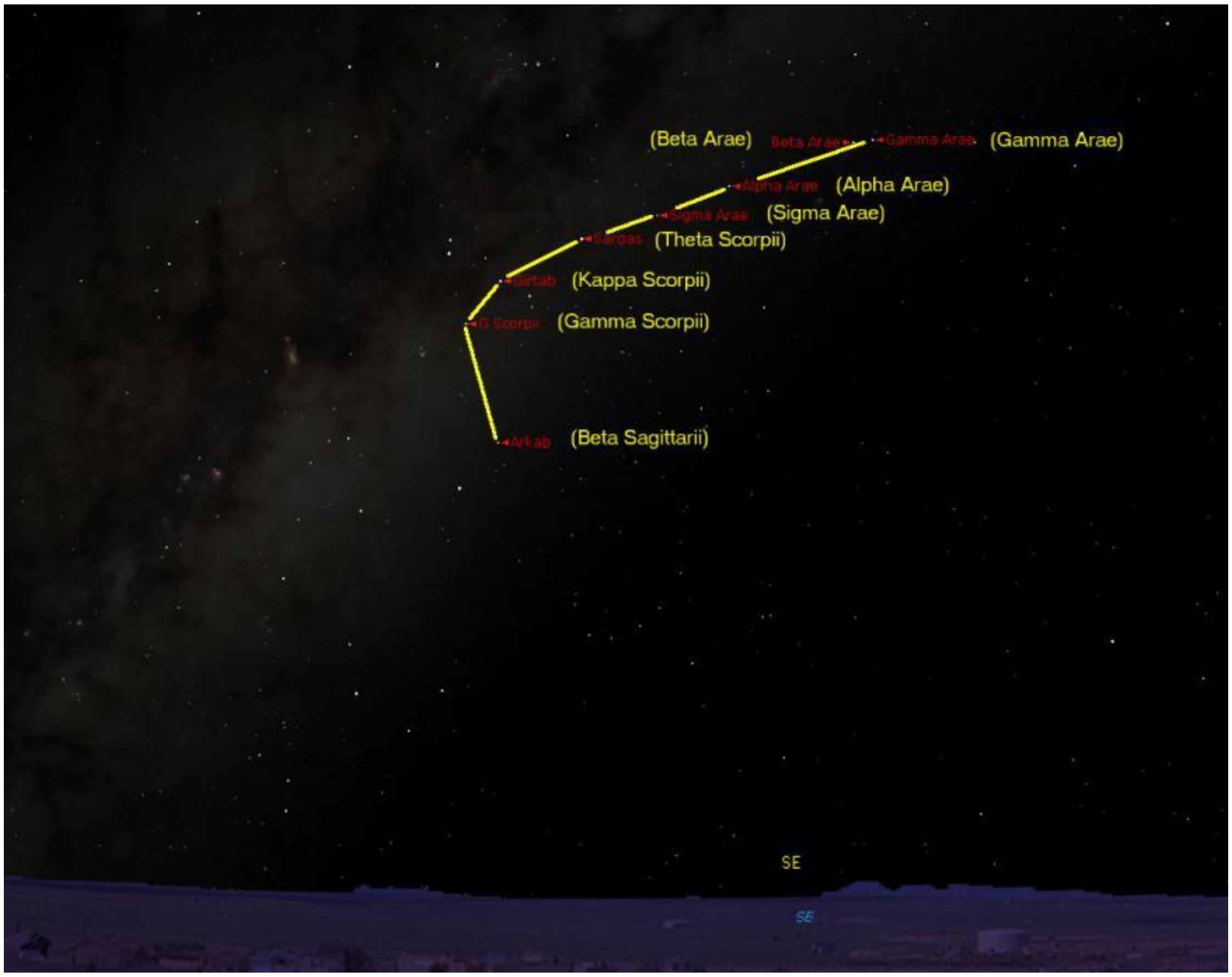

*Fig 3 Goodooga to Bunya Mts. star map (image courtesy of Starry Night Education)*

The same travel is represented on the ground by red lines in Fig 4. The actual ground routes are only loosely similar in direction to the star maps because the star maps do not represent a navigation aid in terms of direction and distance, but just a memory aid to the waypoints. In September, when the travel might commence, the same stars can be seen (higher in the sky and to the southwest), but they have rotated to the point where they would be difficult to use as a navigation aid. It is interesting to note that many main roads (yellow lines in Fig 4) appear to align with the Aboriginal travel routes and some towns appear at the Aboriginal waypoints. According to Kerwin (2010: 159-63) and Norris & Harney (2014), many Aboriginal trading routes became routes of travel for Europeans and stock routes for the movement of animals. These routes later became main roads, and where they turned or split, towns often appeared.

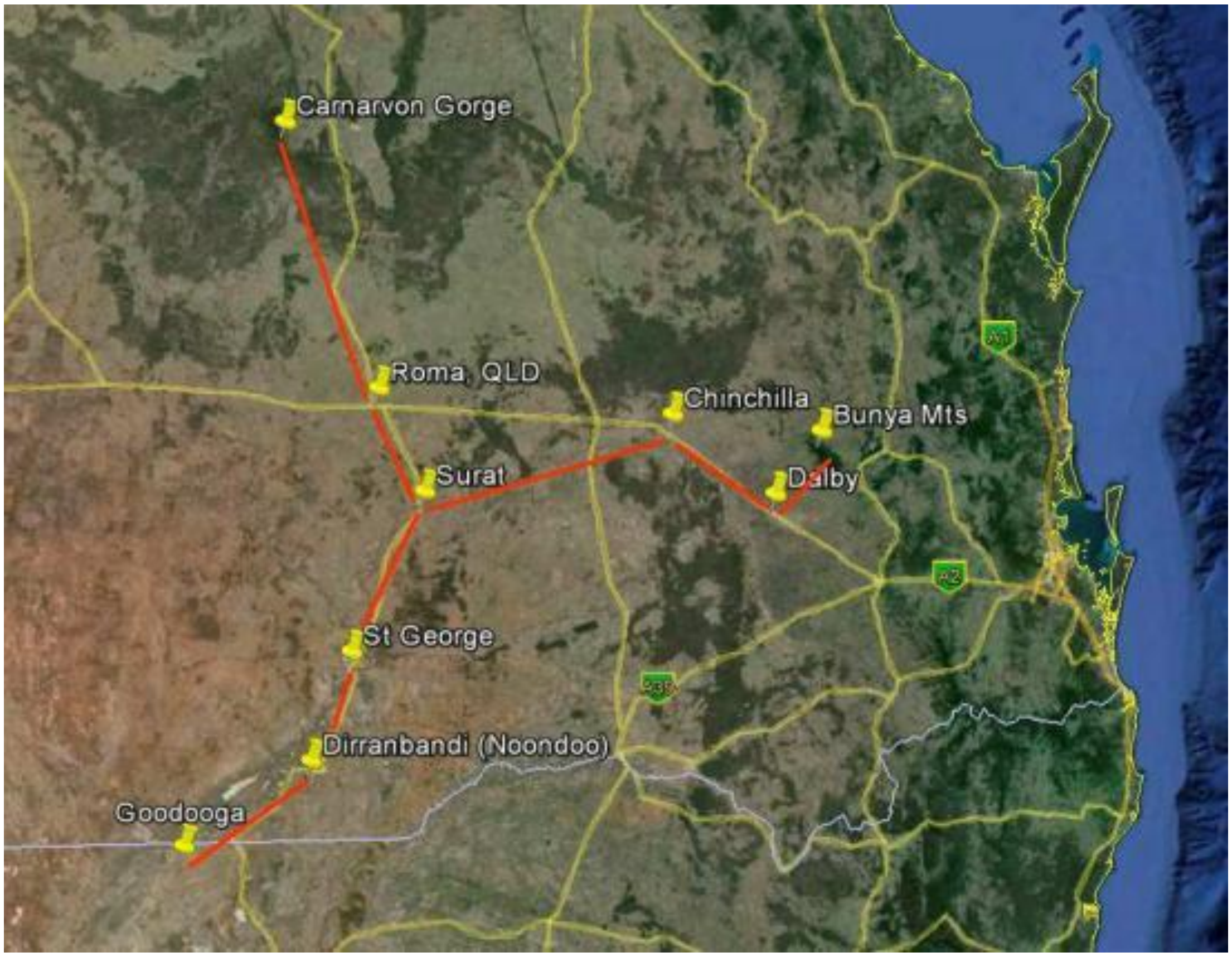

*Fig 4 ground travel routes to Carnarvon Gorge and Bunya Mts. (image Google Earth)*

The concept that these Aboriginal routes were also storylines or Dreaming tracks is reinforced by one of the waypoints on Fig 4, Dirranbandi (Noondoo). This point is actually 25km east of the town of Dirranbandi, at a point called Noondoo Ridge, which is the highest point in the area, and provides a clear view to the south. Anderson has confirmed that the storylines had their beginning at this point because the Southern Cross (Crux) could be clearly seen from this point, and the stories themselves had their origin in the night sky around Crux.

There were other star mapped routes used by the Euahlayi, including one which travelled to the northwest to a waterhole near Quilpie, QLD which was a ceremonial centre for Aboriginal people from a wide area, including the Arrernte people from the Central Desert (a distance of over 900km). The star map started from the winter camp (Beta Sagittarii/Goodooga) to X Sagittarii (Hebel, QLD), to 51 Ophiuchi (Ballon, QLD), to Eta Ophiuchi (Cunnamulla, QLD), to Zeta Ophiuchi (Yowah, QLD), then to Epsilon Ophiuchi (waterhole northwest of Quilpie, QLD). Fig 5 shows the star map, and Fig 6 shows the ground route. The fact that other Aboriginal groups came to such ceremonial meeting points suggests that they also used star maps for teaching travel, but that their star maps were different to those used by the Euahlayi, other than reaching the same point on land. It may also be that the songlines of the Euahlayi routes connected to the songlines of the other groups at the ceremonial places.

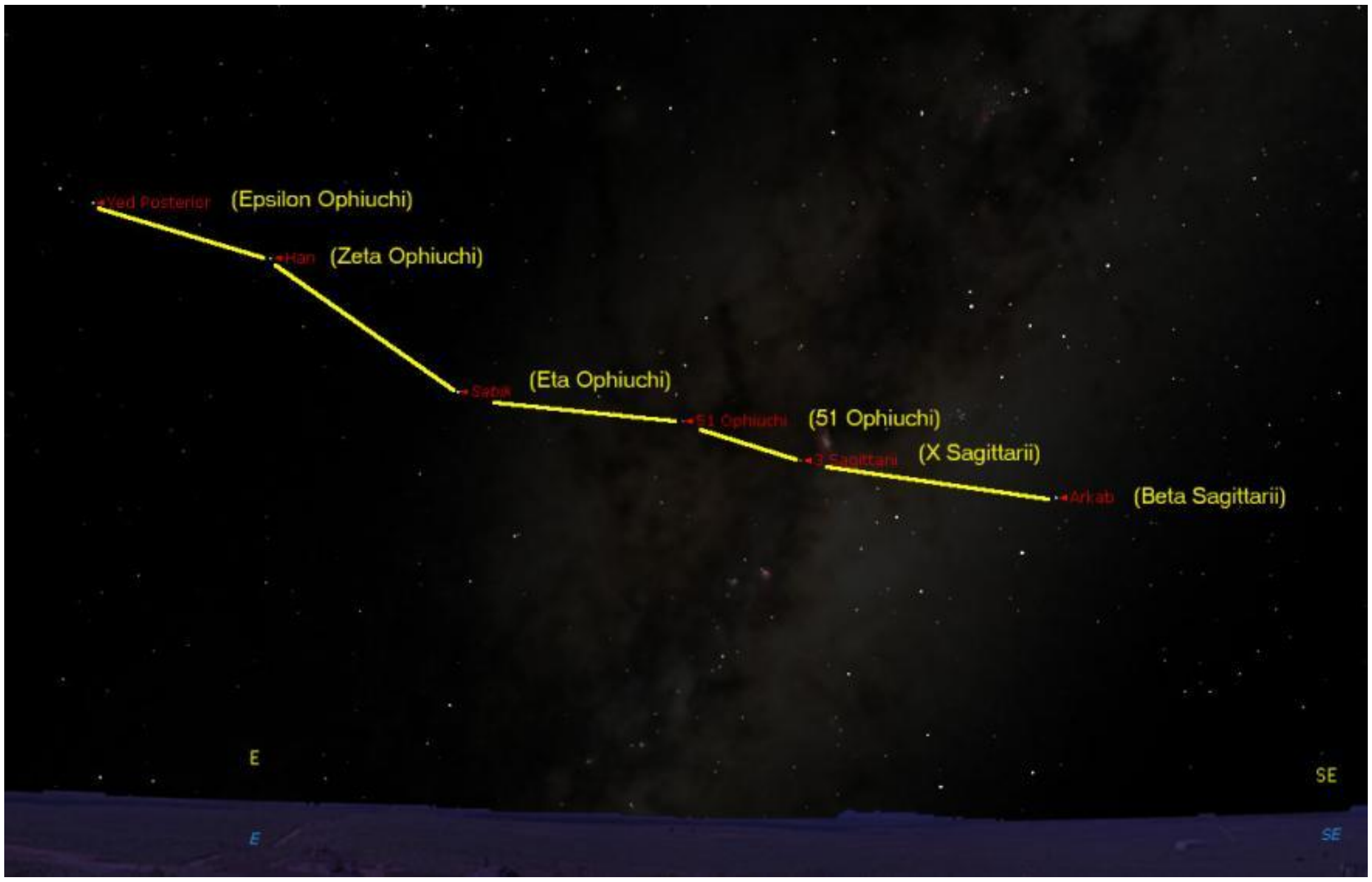

*Fig 5 Goodooga to Quilpie waterhole star map (image courtesy of Starry Night Education)*

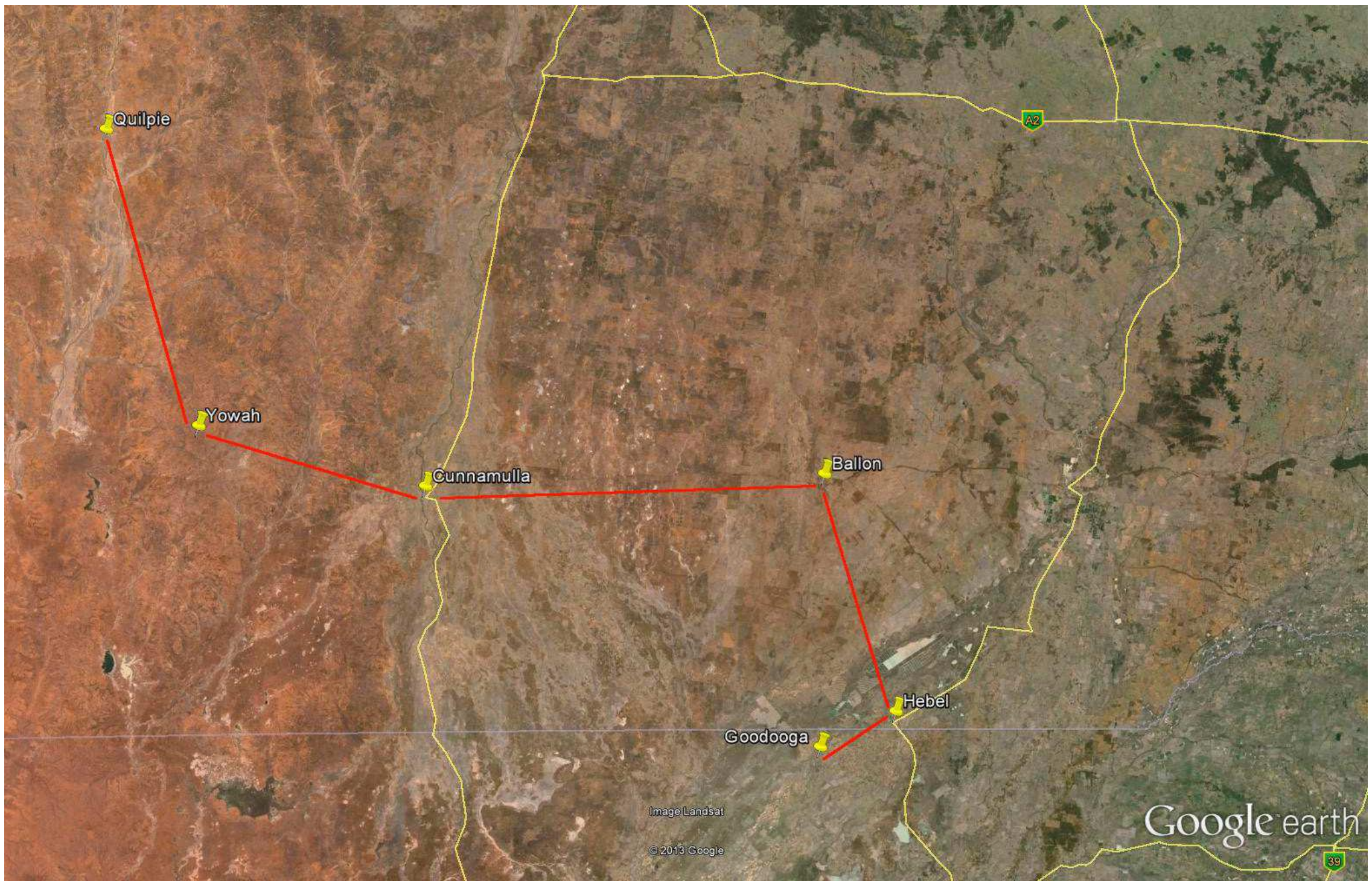

*Fig 6 ground travel route Goodooga to Quilpie waterhole (image Google Earth)*

There were a number of ceremonial sites in QLD and NSW where the Euahlayi people travelled, and the final star map in this study must have been used to educate people about them and their general location. This star map is not a map of waypoints and travel, but simply a representative map of mainly *Bora* grounds. Fig 7 shows the star map, and Fig 8 shows the equivalent ground map.

The Euahlayi, Kamilaroi, and Bigambul people used the *Bora* grounds in this map, but at Sandy Camp a number of language groups joined for ceremonies, including the Euahlayi, Kamilaroi, Murrawarri, Ngemba, Wiradjuri, Wongaibon, and Wailwon.

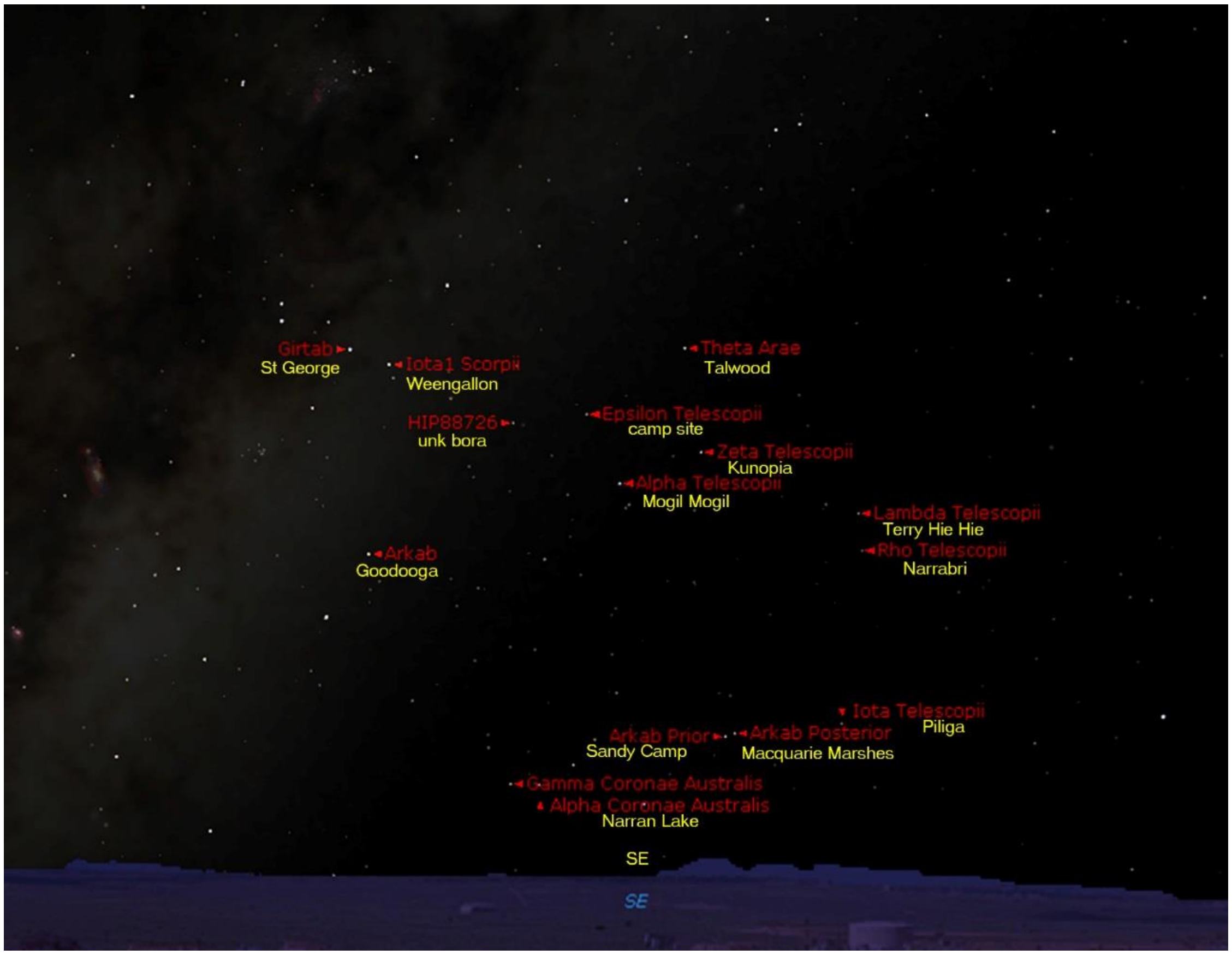

*Fig 7 regional Bora star map (image courtesy of Starry Night Education)*

There is some symmetry between the star map and the ground map if you turn the star map image about 45° to the right, but it was probably not intended to be used as an exact representation, and certain locations, such as Narran Lake, are clearly out of place.

All the stars identified for the star maps described in this study have been checked to see if (1) they are visible in mid-August from north central NSW, and (2) if they are sufficiently bright that they can be seen with the naked eye. In both cases, this was correct. In the second case, from a dark sky location as no doubt existed prior to the light pollution brought by Europeans, the visual limiting magnitude (dimness able to be perceived) is 7.5-8.0 (Bortle, 2001: 126-7). All the stars in the star maps are significantly brighter than the visual limiting magnitude.

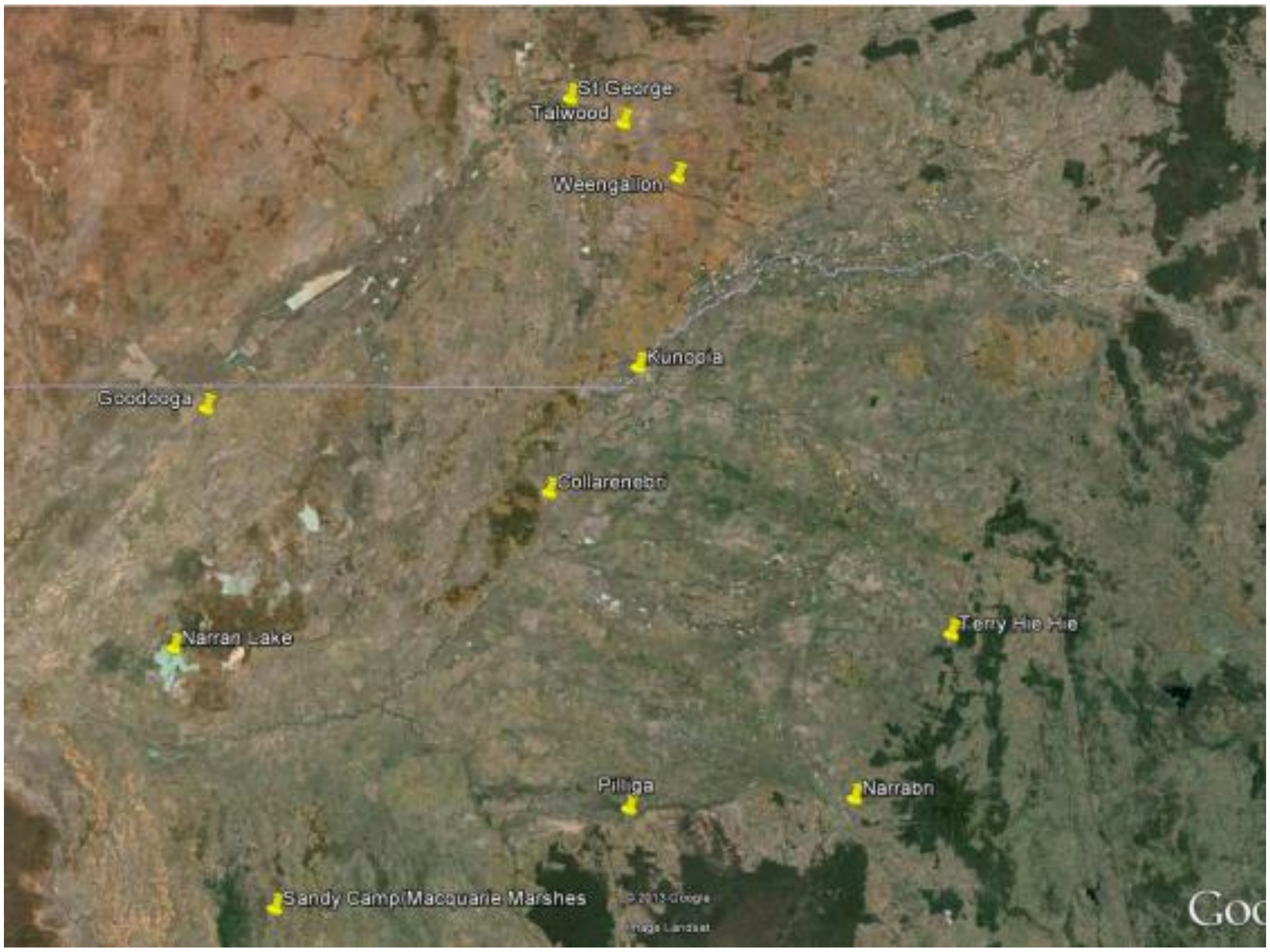

*Fig 8 regional Bora ground map (image Google Earth)*

## 5. Conclusions

Aboriginal people in Australia have a rich and well-developed knowledge of the night sky, which they use in their culture of oral transmission of knowledge, and as a means of assisting in their management of resources. The Euahlayi language group is a relatively small language group located in north-central NSW and south-central QLD, and shares many cultural aspects with the Kamilaroi and other surrounding language groups.

We have shown that while many Aboriginal cultural groups have a rich knowledge of the night sky, this does not necessarily mean that they all use the stars and night sky for actual navigation in the sense of European celestial navigation. This could be because they have no need to travel at night, or could be related to cultural reasons, such as the difference in the use of astronomy between Aboriginal and European peoples.

There is clear evidence that before the European invasion (and after), there was a very well-established and extensive network of trade routes in Australia, used extensively by Aboriginal people for trading in goods, ceremonies and stories, and that these trade routes were aligned with songlines and Dreaming stories, many of which covered vast distances across the Australian continent. These songlines sometimes had their equivalents in the night sky, also crossing great distances, while the connections between the songlines on the ground and in the night sky are yet to be confirmed.

There is no evidence that Aboriginal people possessed portable maps of the night sky, other than some way-finding devices, the purposes of which remain unclear. However, the Euahlayi people used a known pattern of stars in the night sky to teach and remember a number of waypoints on a route to a destination, often a ceremonial gathering place. This star map was used in winter during the planning for the summer travel, and we have identified at least three routes from Euahlayi country to ceremonial or resource destinations. A further star map identified the general location of many of the *Bora* ceremonial grounds in north-central NSW and south-central QLD, but was not used for actual travel. These *Bora* grounds linked at least three language groups in that area, but one ceremonial site was identified as a place for gatherings of up to seven language groups.

Further research on the use of star maps for travel by other language groups, particularly those who might have met the Euahlayi peoples at common ceremonial locations, may lead to a clearer understanding of the Aboriginal use of the night sky for travel.

## Acknowledgements

We acknowledge and pay our respects to the traditional owners and elders, both past and present, of the Kamilaroi, Euahlayi, Ngemba, and Murrawarri peoples. We thank the participants of the Kamilaroi project, Michael Anderson, Rhonda Ashby, Lachie Dennis, Paul Gordon, Greg Griffiths, Brenda McBride, Jason Wilson, and one anonymous person, for their stories and those of their families. We particularly thank Michael Anderson for his detailed story on the star maps, which is central to this study. Michael has acknowledged *Gheedjar* (Walter Sand) and Jack McCrae as his sources for this knowledge, who told him they were taught by *Boobar* (Fred Reece) and Jack Murphy. Anderson was also told that Jack Murphy was taught by a descendant of King Rory from Walgett, who taught Ridley his knowledge of the sky in 1872 (Ridley, 1873: 273-5).